\documentclass[pre,aps,showpacs,twocolumn,protrait]{revtex4}
\usepackage{graphicx,amsmath,bm}
\usepackage{amsfonts}
\usepackage{amssymb}
\usepackage{multirow} 

\begin{document}

\title{Search in weighted complex networks}
\author{Hari P. Thadakamalla and S. R. T. Kumara}
\affiliation{Department of Industrial Engineering, The Pennsylvania
State University, University Park, Pennsylvania, 16802, USA}

\author{R. Albert}
\affiliation{Department of Physics, The Pennsylvania State
University, University Park, Pennsylvania, 16802, USA}

\date{\today}
\begin{abstract}
We study trade-offs presented by local search algorithms in complex
networks which are heterogeneous in edge weights and node degree. We
show that search based on a network measure, local betweenness
centrality (LBC), utilizes the heterogeneity of both node degrees
and edge weights to perform the best in scale-free weighted
networks. The search based on LBC is universal and performs well in
a large class of complex networks.
\end{abstract}
\pacs{89.75.Fb, 89.75.Hc, 02.10.Ox, 89.70.+c}

\maketitle

\section{\label{intro}Introduction}

Many large-scale distributed systems found in communications,
biology or sociology can be represented by complex networks. The
macroscopic properties of these networks have been studied
intensively by the scientific community, which has led to many
significant results ~\cite{Albert2002,Dorogovtsev2002,Newman2003a}.
Graph properties such as the degree distribution and clustering
coefficient were found to be significantly different from random
graphs ~\cite{Erdos1959,Bollabas1985} which are traditionally used
to model these networks. One of the major findings is the presence
of heterogeneity in various properties of the elements in the
network. For instance, a large number of the real-world networks
including the World Wide Web, the Internet, metabolic networks,
phone call graphs, and movie actor collaboration networks are found
to be highly heterogeneous in node degree (i.e., the number of edges
per node) ~\cite{Albert2002,Dorogovtsev2002,Newman2003a}. The
clustering coefficients, quantifying local order and cohesiveness
~\cite{Watts1998}, were also found to be heterogeneous, i.e.,
$C(k)~k^{-1}$ ~\cite{Ravasz2002}. These discoveries along with
others related to the mixing patterns of complex networks initiated
a revival of network modeling in the past few years
~\cite{Albert2002,Dorogovtsev2002,Newman2003a}. Focus has been on
understanding the mechanisms which lead to heterogeneity in node
degree and implications of it on the network properties. It was also
shown that this heterogeneity has a huge impact on the network
properties and processes such as network resilience
~\cite{Albert2000b,Albert2004}, network navigation, local search
~\cite{Adamic2001}, and epidemiological processes
~\cite{Pastor-Satorras2001a,Pastor-Satorras2001b,Pastor-Satorras2002a,Pastor-Satorras2002b,Pastor-Satorras2003}.

Recently, there have been many studies
~\cite{Granovetter1973,Newman2001,Yook2001,Noh2002,Braunstein2003,Barrat2004c,Pimm2002,
Krause2003,Almaas2004,Barrat2004a,Guimera2005,Pastor-Satorras2004,Goh2004}
that tried to analyze and characterize weighted complex networks
where edges are characterized by capacities or strengths instead of
a binary state (present or absent). These studies have shown that
heterogeneity is prevalent in the capacity and strength of the
interconnections in the network as well. Many researchers
~\cite{Granovetter1973,Newman2001,Pimm2002,
Krause2003,Almaas2004,Barrat2004a,Guimera2005,Pastor-Satorras2004,Goh2004}
have pointed out that the diversity of the interaction strengths is
critical in most real-world networks. For instance, sociologists
have shown that the weak links that people have outside their close
circle of friends play a key role in keeping the social system
together ~\cite{Granovetter1973,Newman2001}. The Internet traffic
~\cite{Pastor-Satorras2004} or the number of passengers in the
airline network ~\cite{Barrat2004a,Guimera2005} are critical
dynamical quantities that can be represented by using weighted
edges. Similarly, the diversity of the predator-prey interactions
and of metabolic reactions is considered as a crucial component of
ecosystems ~\cite{Pimm2002} and metabolic networks respectively
~\cite{Krause2003,Almaas2004}. Thus it is incomplete to represent
real-world systems with equal interaction strengths between
different pairs of nodes.

In this paper, we concentrate on finding efficient decentralized
search strategies on networks which have heterogeneity in edge
weights. This is an intriguing and relatively little studied problem
that has many practical applications. Suppose some required
information such as computer files or sensor data is stored at the
nodes of a distributed network. Then to quickly determine the
location of particular information, one should have efficient
decentralized search strategies. This problem has become more
important and relevant due to the advances in technology that led to
many distributed systems such as sensor networks
~\cite{Estrin1999,Raghavan2005}, peer-to-peer networks
~\cite{Kan2001,Hong2001} and dynamic supply chains
~\cite{Thadakamalla2004}. Previous research on local search
algorithms
~\cite{Adamic2001,Kleinberg2000a,Kleinberg2000b,Kleinberg2001,Watts2002,Adamic2003,Arenas2003}
has assumed that all the edges in the network are equivalent. In
this paper we study the complex tradeoffs presented by efficient
local search in weighted complex networks. We simulate and analyze
different search strategies on Erdos-Rényi (ER) random graphs and
scale-free networks. We define a new local parameter called local
betweenness centrality (LBC) and propose a search strategy based on
this parameter. We show that irrespective of the edge weight
distribution this search strategy performs the best in networks with
a power-law degree distribution (i.e., scale-free networks).
Finally, we show that the search strategy based on LBC is usually
equivalent with high-degree search (discussed by Adamic \emph{et}
\emph{al}. ~\cite{Adamic2001}) in un-weighted (binary) networks.
This implies that the search based on LBC is more universal and is
optimal in a larger class of complex networks.

The rest of the paper is organized as follows. In Sec. II, we
describe the problem in detail and briefly discuss the literature
related to search in complex networks. In Sec. III, we define the
local betweenness centrality (LBC) of a node's neighbor and show
that it depends on the weight of the edge connecting the node and
neighbor and on the degree of the neighbor. Section IV explains our
methodology and different search strategies considered. Section V
gives the details of the simulations conducted for comparing these
strategies. In Sec. VI, we discuss the findings from simulations on
ER random and scale-free networks. In Sec. VII, we prove that the
LBC and degree-based search are equivalent in un-weighted networks.
Finally, we give conclusions in Sec. VIII.

\section{\label{Problem}Problem description and literature}

The problem of decentralized search goes back to the famous
experiment by Milgram ~\cite{Milgram1967} illustrating the short
distances in social networks. One of the striking observations of
this study as pointed out by Kleinberg
~\cite{Kleinberg2000a,Kleinberg2000b,Kleinberg2001} was the ability
of the nodes in the network to find short paths by using only local
information. Currently, Watts \emph{et al.} ~\cite{Watts} are doing
an Internet-based study to verify this phenomenon. Kleinberg
demonstrated that the emergence of such phenomenon requires special
topological features
~\cite{Kleinberg2000a,Kleinberg2000b,Kleinberg2001}. Considering a
family of network models that generalizes the Watts-Strogatz model
~\cite{Watts1998}, he showed that only one particular model among
this infinite family can support efficient decentralized algorithms.
Unfortunately, the model given by Kleinberg is too constrained and
represents only a very small subset of complex networks. Watts
\emph{et al.} presented another model to explain the phenomena
observed by Milgram which is based upon plausible hierarchical
social structures ~\cite{Watts2002}. However, in many real-world
networks, it may not be possible to divide the nodes into sets of
groups in a hierarchy depending on the properties of the nodes as in
the Watts \emph{et al.} model.

Recently, Adamic \emph{et al.} ~\cite{Adamic2001} showed that in
networks with a power law degree distribution (scale-free networks)
high degree seeking search is more efficient than random walk
search. In random walk search, the node that has the message passes
it to a randomly chosen neighbor. This process continues untill it
reaches the target node. In high degree search, the node passes the
message to the neighbor that has the highest degree among all nodes
in the neighborhood, assuming that a more connected neighbor has a
higher probability of reaching the target node. The high degree
search was found to outperform the random walk search consistently
in networks having power-law degree distribution for different
exponents varying from 2.0 to 3.0. Using generating function
formalism given by Newman ~\cite{Newman2003b}, Adamic \emph{et al.}
showed that for random walk search the number of steps s until
approximately the whole graph is revealed is given by
$s~N^{3(1-2/\tau)}$, where $\tau$ is the power-law exponent, while
high degree search leads to a much more favorable scaling
$s~N^{2-4/\tau}$.

The assumption of equal edge weights (meaning the cost, bandwidth,
distance, or power consumption associated with the process described
by the edge) usually does not hold in real-world networks. As
pointed out by many researchers
~\cite{Granovetter1973,Newman2001,Yook2001,Noh2002,Braunstein2003,Barrat2004c,Pimm2002,
Krause2003,Almaas2004,Barrat2004a,Guimera2005,Pastor-Satorras2004,Goh2004},
it is incomplete to assume that all the links are equivalent while
studying the dynamics of large-scale networks. The total path length
(\emph{p}) in a weighted network for the path 1 - 2 - 3… - n, is
given by $p = \sum_{i=1}^nw_{i,i+1}$, where $w_{i,i+1}$ is the
weight on the edge from node i to node i+1. Even though high-degree
search results in a path with smaller number of hops, the total path
length may be high if the weights on these edges are high.  Thus, to
be more realistic and closer to real-world networks we need to
explicitly incorporate weights in any proposed search algorithm. In
this paper, we are interested in designing decentralized search
strategies for networks that have the following properties:

\begin{enumerate}
  \item Its node degree distribution follows a power-law in with
exponent varying from 2.0 to 3.0. Although we discuss the search
strategies for networks with Poisson degree distribution (ER random
graphs), we concentrate more on scale free networks since most of
the real world networks are found to exhibit this behavior
~\cite{Albert2002,Dorogovtsev2002,Newman2003a}.
  \item It has non-uniformly distributed weights on the edges. Here
the weights signify the cost/time taken to pass the message/query.
Hence, smaller weights correspond to shorter/better paths. We
consider different distributions such as Beta, uniform, exponential
and power-law.
  \item It is unstructured and decentralized. That is, each node
has information only about its first and second neighbors and no
global information about the target is available. Also, the nodes
can communicate only with their immediate neighbors.
  \item Its topology is dynamic (ad-hoc) while still
maintaining its statistical properties. These particular types of
networks are becoming more prevalent due to advances made in
different areas of engineering especially in sensor networks
~\cite{Estrin1999,Raghavan2005}, peer-to-peer networks
~\cite{Kan2001,Hong2001} and dynamic supply chains
~\cite{Thadakamalla2004}. Here, in this paper we analyze the problem
of finding decentralized algorithms in such weighted complex
networks, which we believe has not been explored to date.
\end{enumerate}

Among the search strategies employed in this paper is a strategy
based on the local betweenness centrality (LBC) of nodes.
Betweenness centrality (also called load), first developed in the
context of social networks ~\cite{Wasserman1994}, has been recently
adapted to optimal transport in weighted complex networks by Goh
\emph{et al.} ~\cite{Goh2004}. These authors have shown that in the
strong disorder limit (that is, when the total path length is
dominated by the maximum edge weight over the path), the load
distribution follows a power-law for both ER random graphs and
scale-free networks. To determine a node's betweenness as defined by
Goh \emph{et al.} one would need to have the knowledge of the entire
network. Here we define a local parameter called local betweenness
centrality (LBC) which only uses information on the first and second
neighbors of a node, and we develop a search strategy based on this
local parameter.

\section{\label{LBC}Local Betweenness Centrality}

\begin{figure}
    \begin{center}
  \includegraphics[width=8.6cm]{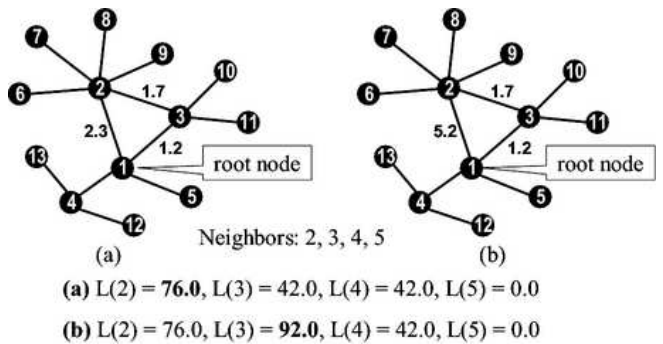}\\
  \caption{(a) In this configuration, neighbor node 2 has a higher LBC than other neighbors 3, 4 and 5.
  This depicts why higher degree for a node helps in obtaining higher LBC.
  (b) However, in this configuration the LBC of the neighbor node 3 is higher than neighbors 2, 4 and 5.
  This is due to the fact that the edge connecting 1 and 2 has a larger weight. These two configurations
  show that the LBC of a neighbor depends both on the edge weight and the node degree. In both cases,
  edge-weights other than those shown in the figure as assumed to be 1.}\label{figure1}
    \end{center}
\end{figure}

We assume that each node in the network has information about its
first and second neighbors. For calculating the local betweenness
centrality of the neighbors of a given node we consider the local
network formed by that node (which we will call the root node), its
first and second neighbors.  Then, the betweenness centrality,
defined as the fraction of shortest paths going through a node
~\cite{Newman2003a}, is calculated for the first neighbors in this
local network. Let $L(i)$ be the LBC of a neighbor node $i$ in the
local network. Then $L(i)$ is given by
\begin{equation*}
    L(i)=\sum_{\stackrel{s \neq n \neq t}{s,t \ \in \ \mathrm{local \
network}}} \frac {\sigma_{st}(i)}{\sigma_{st}},
\end{equation*}
where $\sigma_{st}$ is the total number of shortest paths (where
shortest path means the path over which the sum of weights is
minimal) from node $s$ to $t$. $\sigma_{st}(i)$ is the number of
these shortest paths passing through $i$. If the LBC of a node is
high, it implies that this node is critical in the local network.
Intuitively, we can see that the LBC of a neighbor depends on both
its degree and the weight of the edge connecting it to the root
node. For example, let us consider the networks in Fig.
\ref{figure1}(a) and Fig. \ref{figure1}(b). Suppose that these are
the local networks of node 1. In the network in Fig.
\ref{figure1}(a), node 2 has the highest degree among the neighbors
of node 1 (i.e. nodes 2, 3, 4 and 5). All the shortest paths from
the neighbors of node 2 (6, 7, 8 and 9) to other nodes must have to
pass through node 2. Hence, we see that higher degree for a node
definitely helps in obtaining a higher LBC.

Now consider a similar local network but with a higher weight on the
edge from 2 to 1 as shown in Fig. \ref{figure1}(b). In this network
all the shortest paths through node 2 will also pass through node 3
(2-3-1) instead of going directly from node 2 to node 1. Hence, the
LBC of the neighbor node 3 will be higher than that of neighbor 2.
Thus we clearly see that the LBCs of the neighbors of node 1 depend
on both the neighbors' degrees and the weights on the edges
connecting them. Note that a neighbor having the highest degree or
the smallest weight on the edge connecting it to root node does not
necessarily imply that it will have the highest LBC.

\section{\label{methodology}Methodology}

In un-weighted scale-free networks, Adamic \emph{et}. \emph{al}.
~\cite{Adamic2001} have shown that high degree search which utilizes
the heterogeneity in node degree is efficient. Thus one expects that
in weighted scale-free networks, an efficient search strategy should
consider both the edge weights and node degree. We investigated the
following set of search strategies given in the order of the amount
of information required.

\begin{enumerate}
  \item \emph{Choose a neighbor randomly}: The node tries to reach the target
by passing the message/query to a randomly selected neighbor.
  \item \emph{Choose the neighbor with smallest edge weight}: The node
passes the message along the edge with minimum weight. The idea
behind this strategy is that by choosing a neighbor with minimum
edge weight the expected distance traveled would be less.
  \item \emph{Choose the best-connected neighbor}: The node passes the
message to the neighbor which has the highest degree. The idea here
is that by choosing a neighbor which is well-connected, there is a
higher probability of reaching the target node. Note that this
strategy takes the least number of hops to reach the target
~\cite{Adamic2001}.
  \item Choose the neighbor with the smallest average
weight: The node passes the message to the neighbor which has the
smallest average weight. The average weight of a node is the average
weight of all the edges incident on that node. The idea here is
similar to the second strategy. Instead of passing the message
greedily along the least weighted edge, the algorithm passes to the
node that has the minimum average weight.
  \item Choose the neighbor with the highest LBC: The
node passes the message to the neighbor which has the highest LBC. A
neighbor with highest LBC would imply that many shortest paths in
the local network pass through this neighbor and the node is
critical in the local network. Thus, by passing the message to this
neighbor, the probability of reaching the target node quicker is
higher.
\end{enumerate}

Note that the strategy which depends on LBC utilizes slightly more
information than strategy 4, namely the weights of the edges of the
root node's first neighbors, but it is considerably more
informative: it reflects the heterogeneities in both edge weights
and node degree. Thus we expect that this search will perform better
than the others, that is, it will give smaller path lengths than the
others.

\section{\label{simulations}Simulations}

For comparing the search strategies we used simulations on random
networks with Poisson and power-law degree distributions. For
homogeneous networks we used the Poisson random network model given
by Erd\H{o}s and Rényi ~\cite{Erdos1959}. We considered a network on
$N$ nodes where two nodes are connected with a connection
probability $p$. For scale-free networks, we considered different
values of degree exponent ranging from 2.0 to 3.0 and a degree range
of 2<$k$<$m~N^{1/\tau}$ and generated the network using the method
given by Newman ~\cite{Newman2003b}. Once the network was generated,
we extracted the largest connected component, shown to always exist
for $2< \tau <3.48$ ~\cite{Aiello2000} and in ER networks for
$p>1/N$ ~\cite{Bollabas1985}. We did our analysis on this largest
connected component that contains the majority of the nodes after
verifying that the degree distribution of this largest connected
component is nearly the same as in the original graph. The weights
on the edges were generated from different distributions such as
Beta, uniform, exponential and power-law. We considered these
distributions in the increasing order of their variances to
understand how the heterogeneity in edge weights affects different
search strategies.

Further, we randomly choose K pairs (source and target) of nodes.
The source, and consecutively each node receiving the message, sends
the message to one of its neighbors depending on the search
strategy. The search continues until the message reaches the node
whose neighbor is the target node. In order to avoid passing the
message to a neighbor that has already received it, a list $l_i$ of
all the neighbors that received the message is maintained at each
node $i$. During the search process, if node $i$ passes the message
to its neighbor $j$, which does not have any more neighbors that are
not in the list $l_j$, then the message is routed back to the node
$i$. This particular neighbor $j$ is marked to note that this node
cannot pass the message any further. The average path distance was
calculated for each search strategy from the paths obtained for
these $K$ pairs. We repeated this simulation for 10 to 50 instances
of the Poisson and power-law networks depending on the size of the
network.

\section{\label{analysis}Analysis}

First, we study and compare different search strategies on ER random
graphs. The weights on the edges were generated from an exponential
distribution with mean 5 and variance 25. Table I compares the
performance of each strategy for the networks of size 500, 1000,
1500 and 2000 nodes. We took the connection probability to be $p$ =
.004 and hence a giant connected component always exists
~\cite{Bollabas1985}. From Table I, it is evident that the strategy
which passes the message to the neighbor with the least edge weight
is better than all the other strategies in homogeneous networks.
Remarkably, a search strategy that needs less information than other
strategies (3, 4 and 5), performed best, while high degree search
and LBC did not perform well since the network is highly homogenous
in node degree.

\begin{table*}
  \caption[Comparison of search strategies in a Poisson random network]{Comparison of search strategies in a Poisson random network.
  The edge weights were generated randomly from an exponential distribution with mean 5 and variance 25.
  The values in the table are the average path distances obtained for each search strategy in these networks.
  The strategy which passes the message to the neighbor with the least edge weight performs the best.}\label{table1}
  \centering
  \begin{tabular}{l@{\hspace{0.8cm}}c@{\hspace{0.8cm}}c@{\hspace{0.8cm}}c@{\hspace{0.8cm}}c}
    \\
  \hline \hline
Search strategy &500 nodes &1000 nodes &1500 nodes &2000 nodes \\
                        \\
                        \hline
Random walk &1256.3 &2507.4 &3814.9 &5069.5 \\
Minimum edge weight &\textbf{597.6 }&\textbf{1155.7 } &\textbf{1815.5 } &\textbf{2411.2 }\\
Highest degree  &979.7 &1923.0  &2989.2 &3996.2 \\
Minimum average node weight &832.1 &1652.7 &2540.5 &3368.6 \\
Highest LBC &864.7 &1800.7 &2825.3 &3820.9 \\
  \hline
  \hline
  \end{tabular}
\end{table*}

However, if we decrease the heterogeneity in edge weights (use a
distribution with lesser variance), we observe that high LBC search
performs best (see Table II). In conclusion, when the heterogeneity
of edge weights is high compared to the relative homogeneity of node
degrees, the search strategies which are purely based on edge
weights would perform better. However, as the heterogeneity of the
edge weights decrease the importance of edge weights decreases and
strategies which consider both edge weights and node degree perform
better.

\begin{table*}
  \caption[Comparison of search strategies in a Poisson random network with different edge weight distributions]{Comparison of search strategies in a Poisson random network with 2000 nodes.
  The table gives results for different edge weight distributions. The mean for all the distributions is 5 and variance is $\sigma^2$.
  The values in the table are the average path lengths obtained for each search strategy in these networks.
  When the weight heterogeneity is high, the minimum edge weight search strategy was the best.
  However, when the heterogeneity of edge weights is low, then LBC performs better. }\label{table2}
  \centering
  \begin{tabular}{l@{\hspace{0.8cm}}c@{\hspace{0.8cm}}c@{\hspace{0.8cm}}c@{\hspace{0.8cm}}c}
  \\
  \hline\hline
    &Beta &Uniform &Exp. &Power-law\\
Search strategy &$\sigma^2 = 2.3$ &$\sigma^2 = 8.3$ & $\sigma^2 = 25$  & $\sigma^2 = 4653.8$ \\
                        \hline
Random walk &1271.91 &1284.9  &1253.68 &1479.32 \\
Minimum edge weight &1017.74 &767.405 &577.83 &562.39\\
Highest degree  &994.64 &1014.05 &961.5 &1182.18 \\
Minimum average node weight &1124.48 &954.295 &826.325 &732.93 \\
Highest LBC &980.65 &968.775 &900.365 &908.48 \\

  \hline
  \hline
  \end{tabular}
\end{table*}

\begin{figure}
    \begin{center}
  \includegraphics[width=8.6cm]{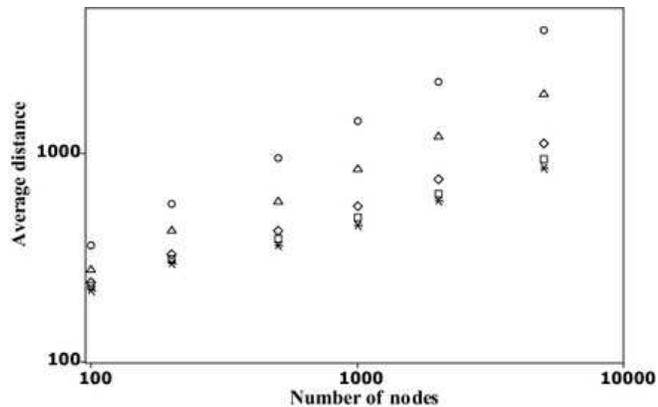}\\
  \caption{Scaling for search strategies in power-law networks with exponent 2.1.
            The edge weights are generated from an exponential distribution with
            mean 10 and variance 100. The symbols represent random walk
            ($\circ$) and search algorithms based on minimum edge weight
            ($\square$), high degree ($\diamond$), minimum average node weight
            ($\vartriangle$) and high LBC ($\ast$).}\label{figure2}
    \end{center}
\end{figure}

Next we investigated how the search strategies perform on scale-free
networks. Figure \ref{figure2} shows the scaling of different search
strategies for power-law networks with exponent 2.1. As conjectured,
the search strategy that utilizes the heterogeneities of both the
edge weights and nodes' degrees (the high LBC search) performed
better than the other strategies. A similar phenomenon was observed
for different exponents of the power-law network (see Table III).
Except for the power-law exponent 2.9, the high LBC search was
consistently better than others. We observe that as the
heterogeneity in the node degree decreases (i.e. as power-law
exponent increases), the difference between the high LBC search and
other strategies decreases. When the exponent is 2.9, the
performance of LBC, minimum edge weight and high degree searches
were almost the same. Note that when the network becomes homogeneous
in node degree the minimum edge weight search performs better than
high LBC search (Table I). This implies that similarly to high
degree search ~\cite{Adamic2001}, the effectiveness of high LBC
search also depends on the heterogeneity in node degree.

\begin{table*}
  \caption[Comparison of search strategies in power-law network with different power-law exponents]{Comparison of search strategies in power-law network on 2000 nodes with different power-law exponents.
  The edge weights are generated from an exponential distribution with mean 5 and variance 25.
  The values in the table are the average path lengths obtained for each search strategy in these networks.
  LBC search, which reflects both the heterogeneities in edge weights and node degree, performed the best for all power-law exponents.
  The systematic increase in all path lengths with the increase of the power-law exponent $\tau$
   is due to the fact that the average degree of the network decreases with $\tau$.}\label{table3}
  \centering
  \begin{tabular}{l@{\hspace{0.8cm}}c@{\hspace{0.8cm}}c@{\hspace{0.8cm}}c@{\hspace{0.8cm}}c@{\hspace{0.8cm}}c}
  \\
  \hline\hline
power-law exponent $=$ &2.1 &2.3 &2.5 &2.7 &2.9 \\
Search strategy \\
                        \hline
Random walk &1108.70 &1760.58 &2713.11 &3894.91 &4769.75\\
Minimum edge weight &318.95 &745.41 &1539.23 &2732.01 &3789.56 \\
Highest degree  &375.83 &761.45 &1519.74 &2693.62 &3739.61  \\
Minimum average node weight &605.41 &1065.34 &1870.43 &3042.27 &3936.03 \\
Highest LBC &\textbf{298.06} &\textbf{707.25} &\textbf{1490.48} &\textbf{2667.74} &\textbf{3751.53} \\
  \hline
  \hline
  \end{tabular}
\end{table*}

Table IV shows the performance of all the strategies on a scale-free
network (exponent 2.1) with different edge weight distributions. The
percentage values in the brackets show by how much the average
distance for that search is higher than the average distance
obtained by the high LBC search. As in random graphs, we observe
that the impact of edge weights on search strategies increases as
the heterogeneity of the edge weights increase. For instance, when
the variance (heterogeneity) of edge weights is small, high degree
search is better than the minimum edge weight search. On the other
hand, when the variance (heterogeneity) of edge-weights is high, the
minimum edge weight strategy is better than high degree search. In
each case, the high LBC search which reflects both edge weights and
node degree always out-performed the other strategies. Thus, it is
clear that in power-law networks, irrespective of the edge weight
distribution and the power-law exponent, high LBC search always
performs better than the other strategies (Tables III and IV).

\begin{table*}
  \caption[Comparison of search strategies in a power-law network with different edge weight distributions]{Comparison of different search strategies in power-law networks with exponent 2.1 and 2000 nodes with different edge weight
  distributions. The mean for all the edge weight distributions is 5 and the variance is $\sigma^2$. The values in the table are the average
  distances obtained for each search strategy in these networks. The values in the brackets show the relative difference between
  average distance for each strategy with respect to the average distance obtained by the LBC strategy. LBC search,
  which reflects both the heterogeneities in edge weights and node degree, performed the best for all edge weight distributions.}\label{table4}

  \centering
  \begin{tabular}{l@{\hspace{0.8cm}}c@{\hspace{0.8cm}}c@{\hspace{0.8cm}}c@{\hspace{0.8cm}}c}
    \\
  \hline\hline
    &Beta &Uniform &Exp. &Power-law\\
Search strategy &$\sigma^2 = 2.3$ &$\sigma^2 = 8.3$ & $\sigma^2 = 25$  & $\sigma^2 = 4653.8$ \\
                        \hline
\multirow{2}{*}{Random walk} &1107.71 &1097.72 &1108.70 &1011.21 \\
 &(202\%) &(241\%) &(272\%) &(344\%)\\
 \hline
\multirow{2}{*}{Minimum edge weight} &704.47 &414.71 &318.95 &358.54 \\
& (92\%) &(29\%) &(7\%) &(44\%)\\
\hline
\multirow{2}{*}{Highest degree} &379.98 &368.43 &375.83 &394.99 \\
&(4\%) &(14\%) &(26\%) &(59\%)\\
\hline
\multirow{2}{*}{Minimum average node weight} &1228.68 &788.15 &605.41 &466.18 \\
& (235\%) &(145\%) &(103\%) &(88\%) \\
\hline
Highest LBC &\textbf{366.26}  &\textbf{322.30}  &\textbf{298.06}  &\textbf{247.77}\\

  \hline
  \hline
  \end{tabular}
\end{table*}

\begin{figure}
    \begin{center}
  \includegraphics[width=8.6cm]{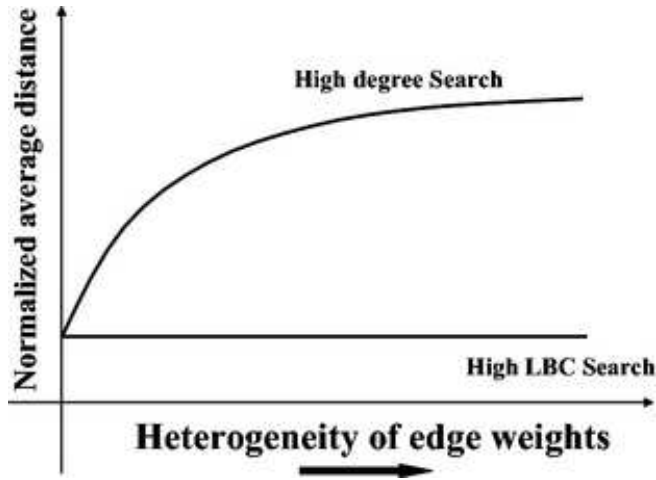}\\
  \caption{The pictorial comparison of the behavior of high degree and high LBC search as the heterogeneity of edge weights
            increases in scale-free networks. Note that average distances are normalized with respect to high
            LBC search.}\label{figure3}
    \end{center}
\end{figure}

Figure \ref{figure3} gives a pictorial comparison of the behavior of
high degree and high LBC search as the heterogeneity of the edge
weights increase (based on the results shown in Table IV). Since
many studies
~\cite{Granovetter1973,Newman2001,Yook2001,Noh2002,Braunstein2003,Barrat2004c,Pimm2002,
Krause2003,Almaas2004,Barrat2004a,Guimera2005,Pastor-Satorras2004,Goh2004}
have shown that there is a large heterogeneity in the capacity and
strengths of the interconnections in the real networks, it is
important that local search is based on LBC rather than high degree
as shown by Adamic \emph{et}. \emph{al}. ~\cite{Adamic2001}.

Note that LBC has been adopted from the definition of betweenness
centrality (BC) which requires the global knowledge of the network.
BC is defined as the fraction of shortest paths among all nodes in
the network that pass through a given node and measures how critical
the node is for optimal transport in complex networks. In
un-weighted scale-free networks there exists a scaling relation
between node betweenness centrality and  degree, $BC~k^\eta$
~\cite{Goh2001}. This implies that the higher the degree, the higher
is the BC of the node. This may be the reason why high degree search
is optimal in un-weighted scale-free networks (as shown by Adamic
\emph{et} \emph{al}. ~\cite{Adamic2001}). However, Goh \emph{et}
\emph{al}. ~\cite{Goh2004} have shown that no scaling relation
exists between node degree and betweenness centrality in weighted
complex networks. It will be interesting to see the relationship
between local and global betweenness centrality in our future work.
Also, note that the minimum average node weight strategy (strategy
4) uses slightly less information than LBC search. However, LBC
search consistently and significantly outperforms it (see Tables I,
II, III and IV). This implies that LBC search uses the information
correctly.

\section{\label{lbconunweighted}LBC on Unweighted Networks}

\begin{figure}
    \begin{center}
  \includegraphics[width= 8.4cm]{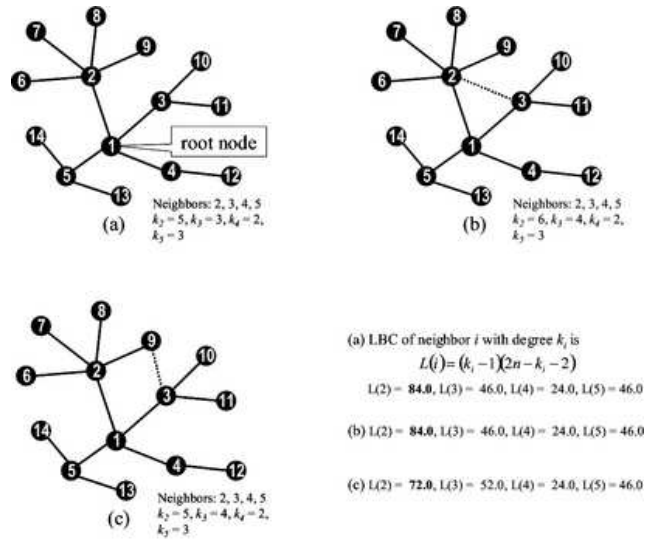}\\
  \caption[LBC in un-weighted networks]{(a) A configuration of a local network with a tree like structure.
            In such local networks, the neighbor with the highest degree has the highest LBC.
            (b) A local network with an edge between two first neighbors. Here again the neighbor with the highest
            degree has the highest LBC. (c) A local network with an edge between a first neighbor and a second neighbor.
            Although there is change in LBCs of neighbors, the order remains the same.}\label{figure4}
    \end{center}
\end{figure}

In this section, we show that the neighbor with the highest LBC is
usually the same as the neighbor with the highest degree in
unweighted networks. Hence, high LBC search would give identical
results as high degree search in un-weighted networks. As mentioned
earlier, in unweighted networks, there is a scaling relation between
the (global) BC of a node and its degree, as $BC\sim k^\eta$
\cite{Goh2001}. However, this does not imply that in an unweighted
local network the neighbor with highest LBC is always the same as
the neighbor with the highest degree. Here, we show that in most
cases the highest degree and the highest LBC neighbors coincide.
First, let us consider a tree-like local network without any loops
similar to the network configuration shown in figure
\ref{figure4}(a). In a local network, there are three types of
nodes, namely, root node, first neighbors and second neighbors. Let
the degree of the root node be $d$ and the degree of the neighbors
be $k_1, k_2, k_3 ... k_d$. The number of nodes ($n$) in the local
network is $n=1+\sum_{j=1}^dk_j$ [one root node, $d$ first neighbors
and $\sum_{j=1}^d(k_j-1)$ second neighbors]. In a tree network there
is a single shortest path between any pair of nodes $s$ and $t$,
thus $\sigma_{st}(i)$ is either zero or one. Then the LBC of a first
neighbor $i$ is given by $L(i)=(k_i-1)(n-2)+(k_i-1)(n-k_i)$ where
$k_i$ is the degree of the neighbor. The first term is due to the
shortest paths from $k_i-1$ neighbors of node $i$ to $n-2$ remaining
nodes (other than node $i$ and the neighbor $j$) in the network. The
second term is due to the shortest paths from $n-k_i$ nodes (other
than $k_i-1$ neighbors and node $i$) to $k_i-1$ neighbors of node
$i$. Note that we chose not to explicitly take into account of the
symmetry of distance in undirected networks and count the s-t and
t-s paths separately. $L(i)$ is an increasing function if
$k_i<n-\frac{1}{2}$, a condition that is always satisfied since
$n=1+\sum_{j=1}^dk_j$ . This implies that in a local network with
tree-like structure, the neighbor with highest degree has the
highest LBC. We extend the above result for other configurations of
the local network by considering different possible cases.

The possible edges other than the edges present in a tree-like local
network are an edge between two first neighbors, an edge between a
first neighbor and a second neighbor and an edge between two second
neighbors. As shown in figure \ref{figure4}(b), an edge among two
first neighbors changes the LBC of the root node but not that of the
neighbors. Figure \ref{figure4}(c) shows a configuration of a local
network with an edge added between a first and a second neighbor.
Now, there is a small change in the LBCs of the neighbors (nodes 2
and 3) which are connected to a common second neighbor (node 9).
Since node 9 is now shared by neighbors 2 and 3, the LBC contributed
by node 9 is divided between these two neighbors. The LBC of such a
neighbor i is $L(i)=(k_i-2)(n-2)+(k_i-2)(n-k_i)+(n-k_j-1)$ where
$k_i$ is the degree of the neighbor $i$ and $k_j$ is the degree of
the neighbor with which node $i$ has a common second neighbor. The
decrease in the LBC of neighbor $i$ is $(n-k_i+k_j-1)$. If there are
two neighbors with the same degree (one with a common second
neighbor and another without any) then the neighbor without any
common second neighbors will have higher LBC. Another possible
change of order with respect to LBC would be with a neighbor $l$ of
degree $k_l = k_i-1$ (if it exists). However,
$L(i)-L(l)=(n-k_i-k_j+1)$ is always greater than 0, since
$n=\sum_{j=1}^dk_j$ in this local network. Thus the only scenario
under which the order of neighbors with respect to LBC is different
than their order with respect to degree when adding an edge between
first and second neighbors is if that creates two first neighbors
with the same degree. A similar argument leads to an identical
conclusion in the case of adding an edge between two second
neighbors as well.

\begin{figure}
    \begin{center}
  \includegraphics[width=8.4cm]{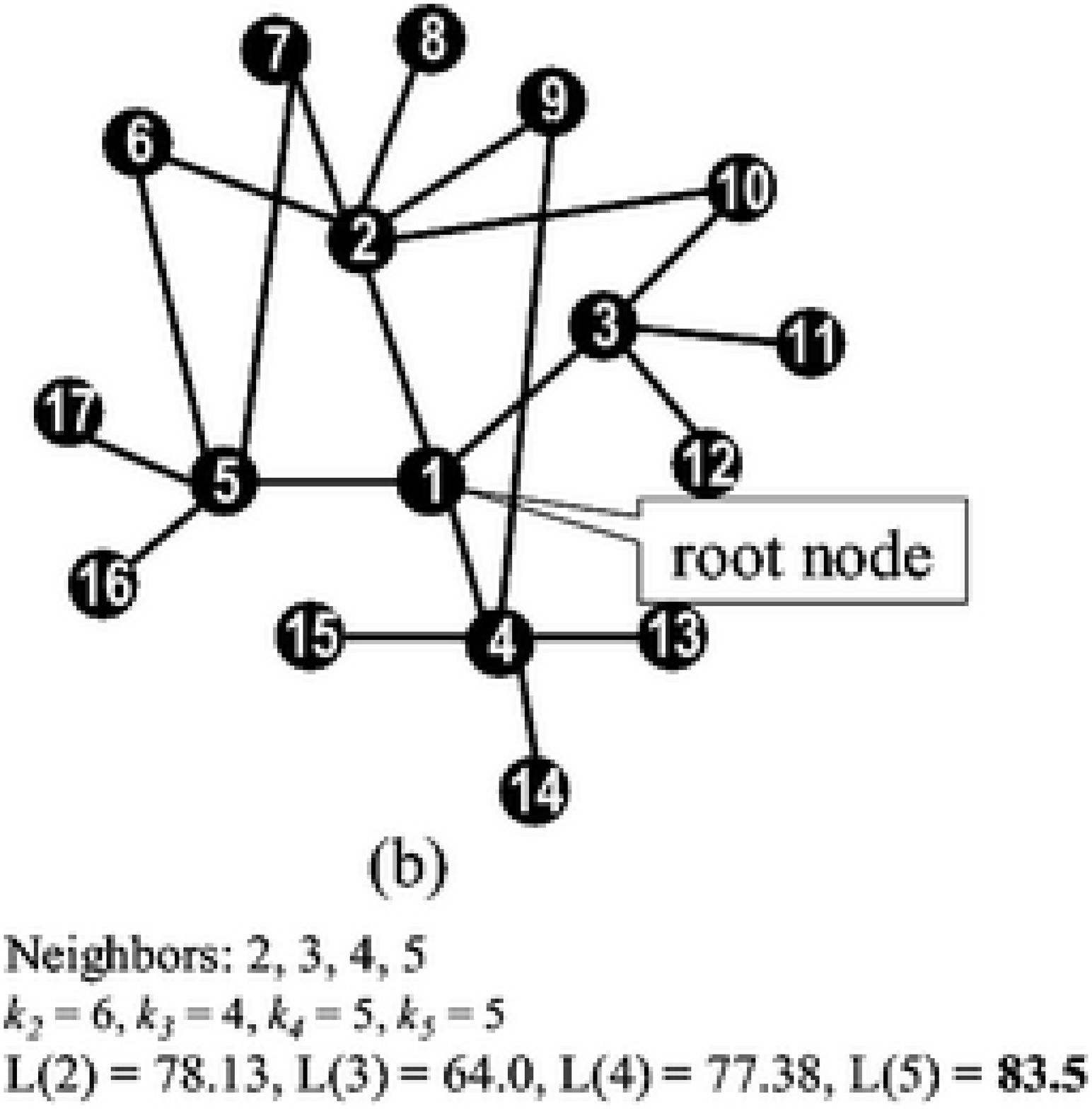}\\
  \caption[Exception for LBC in un-weighted networks]{An instance of a local network where the order of neighbors with respect to LBC is not
        same as the order with respect to node degree.}\label{figure5}
    \end{center}
\end{figure}

The above discussion suggests that the highest degree neighbor is
always the same as the highest LBC neighbor. This is not true in few
peculiar instances of local networks. For example, consider the
network shown in figure \ref{figure5} which has several edges
between the first and second neighbors. We see that the highest
degree neighbor is not the same as the highest LBC neighbor. In this
local network, the highest degree first neighbor (node 2),
participates in several four-node circuits that include the root
node. Thus, there are multiple shortest paths starting from
second-neighbor nodes on these cycles (nodes 6, 7, 9, 10) and the
contributions to node 2's LBC from the paths that pass through it
are smaller than unity, consequently the LBC of node 2 will be
relatively small. This may be one of the reasons why the
highest-degree neighbor node 2 is not the highest LBC neighbor. We
feel that this happens only in some special instances of local
networks. From about 50,000 simulations we found that in 99.63 \% of
cases the highest degree neighbor is the same as the highest LBC
neighbor. Hence, we can conclude that in un-weighted networks the
neighbor with highest LBC is usually identical to the neighbor with
the highest degree.

\begin{acknowledgments}
The authors would like to acknowledge the National Science
Foundation (Grant \# SST 0427840) and a Sloan Research Fellowship to
one of the authors (R. A.) for making this work feasible. Any
opinions, findings and conclusions or recommendations expressed in
this material are those of the author(s) and do not necessarily
reflect the views of the National Science Foundation (NSF). In
addition, the first author (H.P.T.) would like to thank Usha Nandini
Raghavan for interesting discussions on issues related to this work.
\end{acknowledgments}

\bibliographystyle{apsrev}
\bibliography{Biblio-Database}

\end{document}